\documentclass[prd, aps, nobibnotes, nofootinbib,
eqsecnum,
preprint,
preprintnumbers,
amsmath,
amssymb]{revtex4}

\usepackage{graphicx}
\usepackage{bm}
\usepackage{hyperref}
\usepackage{mathrsfs}
\usepackage{xcolor}
\usepackage{multirow}
\usepackage{hhline}

\def\Tr{\mathrm{Tr}}

\begin{document}
	
	\title{{\bf{\Large Cosmology with modified continuity equation in asymptotically safe gravity }}}

	\author{ Rituparna Mandal}
	\email{drimit.ritu@gmail.com, rituparna1992@bose.res.in}\,
	\affiliation{Department of Theoretical Sciences, S.N. Bose National Centre for Basic Sciences, Block JD, Sector III, Salt Lake, Kolkata 700106, India}
	\author{ Sunandan Gangopadhyay}
	\email{sunandan.gangopadhyay@gmail.com, sunandan.gangopadhyay@bose.res.in}
	\affiliation{Department of Theoretical Sciences, S.N. Bose National Centre for Basic Sciences, Block JD, Sector III, Salt Lake, Kolkata 700106, India}
	\author{ Amitabha Lahiri}
	\email{amitabha@bose.res.in }
	\affiliation{Department of Theoretical Sciences, S.N. Bose National Centre for Basic Sciences, Block JD, Sector III, Salt Lake, Kolkata 700106, India}

	\date{\today}

\begin{abstract}
We study FLRW cosmology, taking into account quantum gravitational corrections in the formalism of the exact renormalization group flow of the effective action for gravity.  We calculate the quantum corrected scale factor, energy density, and entropy production at late times, taking different cut-off functions. Our approach differs from previous ones in the way energy-momentum conservation is imposed -- we include the running Newton constant $G(k)$ in the definition of energy-momentum tensor, keeping in mind the covariant conservation identity of the Einstein tensor. The quantum corrections obtained in this approach are different from what are found by letting the conservation equation remain the same as for a scale-independent Newton constant. We also find that for a specific choice of the cut-off scale, the quantum corrected behaviour of the Newton constant and the cosmological constant leads to a bouncing emergent universe solution.
\end{abstract}
\vskip 1cm

\maketitle
	
\section{Introduction}
A consistent quantization of general relativity in four dimensions
is the holy grail of theoretical high-energy and gravitational physics. 
Perturbative renormalization of Einstein gravity fails because of the dimensionful nature of the gravitational coupling.
	An alternative view is that general relativity cannot be quantized directly, but emerges as an effective
	low energy theory from a quantum action which in principle includes all diffeomorphism-invariant local
	functions of the metric not ruled out by other symmetries~\cite{Weinberg:1980gg}. Over the last couple of decades 
	a new approach to quantum gravity based on this view has become popular.
	This is the functional renormalization group (RG) approach for gravity in which a scale-dependent
	RG equation is derived by including all possible diffeomorphism invariant local functions 
	of the metric~\cite{Reuter:2001ag, Wetterich:1992yh,Reuter:1993kw, Niedermaier:2009ep, Reuter:1996cp, Lauscher:2002sq, odi1, odig}. 
	The low energy effective action is obtained by solving the exact functional RG flow equation 
	in terms of a momentum scale parameter $k$. This results in a running Newton's constant as well as 
	running cosmological constant. Einstein equation is then ``improved'' by replacing the Newton's 
	constant and cosmological constant by their scale-dependent versions. The theory 
	that comes from this formulation has its bare action corresponding to a non-trivial fixed point of the RG flow. 
	The existence of a UV fixed point also indicates that the theory is asymptotically safe~\cite{Percacci:2005wu, 
		Niedermaier:2006wt, Niedermaier:2006ns, Niedermaier:2010zz, Nink:2012vd, mniedermair:2003, Litim:2014uca, Falls:2014tra}.

	We use a formulation of asymptotically safe gravity based on an (Euclidean) ``effective average action'' $\Gamma_{k} [g_{\mu \nu}]$~\cite{Reuter:1996cp, Reuter:2001ag, Wetterich:1992yh,Reuter:1993kw, Niedermaier:2006ns, reuter:2012qeg, 0lauscher:2007, Codello:2008vh, Eichhorn:2018, Souma:1999at, litim:2004ie, damel:2014, Manrique:2008zw, Lauscher:2002ijmp, Reuter:2006ijmp, Reuter:2019book, lit2}. It is defined such that it correctly describes all gravitational phenomena, taking into account the effect of all loops, at a momentum scale $k$\,. The effective average action is based on a modified theory whose action includes, in addition to the bare action of the theory, a regulator term which suppresses all IR modes with momenta $p^2< k^2$ below a chosen cutoff $k$\,.
	This results in the exclusion of the modes with $p^2 < k^2$, while the modes
	with $p^2 > k^2$ are integrated out. As a result of this $\Gamma_{k}$ interpolates between the classical action $S= \Gamma_{k\to \infty}$ that corresponds to ignoring all quantum modes, and the standard quantum effective action 
	$\Gamma= \Gamma_{k=0}$ that corresponds to the removal of the IR cutoff. Hence $\Gamma_{k}$ describes a trajectory satisfying a renormalization group flow equation. 
	In order to solve the RG flow equation, a method of  ``truncation in theory space'' is used to keep only the $\sqrt{g}$ and $\sqrt{g}R$ terms in the action. Then the truncated effective average action has the form
	\begin{eqnarray}
		\Gamma_{k}[g,\bar{g}]= \frac{1}{16\pi G(k)} \int d^{d}x \sqrt{g}\left\lbrace -R(g)+2{\Lambda}(k)\right\rbrace + S_{gf}[g,\bar{g}]\,.
		\label{EA}
	\end{eqnarray}
	The RG flow equation obtained using this leads to scale dependent expressions for Newton's gravitational constant and the cosmological constant, 
	\begin{align}
		G(k) &= G _{0} \left [ 1-\omega  G_{0} k^{2} +\mathcal{O}(G^{2}_{0}k^{4}) \right ]\,, \label{G.intro} \\ 
		\Lambda(k) & = \nu G_{0} k^{4}  +\mathcal{O}(G^{2}_{0}k^{6})\,
		\label{flamda.intro}
	\end{align}
	where $G_0$ is the Newton's gravitational constant at $k=0$. 
	The constants appearing in Einstein's field equation are then replaced by the running constants to get the ``RG-improved'' Einstein equation.
	An outline of obtaining the running Newton's gravitational constant and running cosmological constant is presented in an Appendix.

	
	The Friedmann-Lema\^ itre-Robertson-Walker (FLRW) model of cosmology was investigated 
	in~\cite{Bonanno:2001xi, Grande:2011, Cai:2011, Reuter:2005jcap, Bonnano:2012njp, Bonnano:2002plb} 
	using the ``RG improved'' Einstein equation. Bianchi I cosmology 
	was also studied recently in~\cite{Mandal:2019xlg} in the same approach. The procedure leads 
	to a set of ordinary coupled differential equations involving the scale factor, density, 
	Newton's universal gravitational constant, cosmological constant, and a cutoff function.
	An important input that goes into the derivation of the coupled differential equations 
	for the scale factor and density is the covariant conservation of the energy-momentum tensor. 
	Since the Einstein tensor is divergence-free as a geometrical identity, the conservation 
	of the energy-momentum tensor leads to an additional equation involving the time derivatives 
	of the Newton's constant and the cosmological constant. 
	This leads to a consistency equation when we assume an equation of state for the matter;
	once the equation of state is used, we find a system of three equations and two unknowns. 
	The solution of the two unknowns from two of the equations would necessarily need to 
	satisfy the third equation for consistency of the solution. 
	
	If the matter energy-momentum tensor corresponds to a classical perfect fluid, one may 
	be able to justify the use of the usual conservation equation and thus of the consistency
	condition. If however there is an underlying quantum theory of the matter, we should in 
	principle include the RG flow of that as well. For example, if electromagnetic radiation 
	appears as a source in the RG improved Einstein equations, its RG flow should also be 
	considered, resulting in the running fine structure constant appearing in Einstein 
	equation. In that case it turns out that if we demand that the energy-momentum tensor remain
	divergence-free, the resulting consistency equation cannot be satisfied at late
	times~\cite{inprep}. Therefore we should demand conservation, not of the usual $T_{\mu\nu}$, 
	but of the entire right hand side of Einstein equation including the running gravitational 
	and other coupling constants as 
	well as the running cosmological constant. We call this the ``modified conservation equation''
	and apply it to the FLRW cosmology in this paper. We will consider only perfect fluid here as 
	we are mainly interested in finding out 
	how the resulting cosmology differs at late times from the other approach; the 
	problem of including radiation, and thus the running fine structure constant, will 
	be taken up elsewhere. 
		
	The organization of this paper is as follows. In Sec.~\ref{sflrw}, we write the RG improved Einstein equation for FLRW cosmology.
	In Sec.~\ref{Hentropy} we explain our procedure for a particular choice of the relation between the momentum cutoff scale and cosmological time and present our main results for the scale factor and the entropy production rate. In Sec.~\ref{scalea}, we consider some other choices for the momentum cutoff scale. A summary of the results is given in the last section. An outline of how the effective average action functional leads to the flow equation for the scale dependent gravitational constant $G(k)$ and cosmological constant $\Lambda(k)$ is given in an Appendix.

	\section{Improved Field Equations and the cutoff identication}\label{sflrw}
	The idea of using the modified conservation law to study cosmology in the context of scale-dependent 
	(or time-dependent) Newton's constant is fairly general and should be widely applicable, even outside 
	the context of asymptotically safe gravity, and for any kind of cosmological metric. In this paper we 
	will restrict ourselves and apply the modified conservation law to only the spatially flat FLRW metric, 
	\begin{eqnarray}
		ds^2=-dt^{2}+a^{2}(t) \left[dr^{2}+r^{2}\left(d\theta^{2}+\sin^{2}\theta d\phi^{2}\right)\right]\,.
		\label{flrw}
	\end{eqnarray}
	We will take the cosmic matter to be a perfect fluid, for which the energy-momentum tensor reads
	\begin{eqnarray}
		T_{\mu \nu}=(p+\rho)v_{\mu}v_{\nu}+p g_{\mu \nu}\,,
		\label{fluid-emtensor}
	\end{eqnarray}
	where $p$ is the pressure, $\rho$ is the energy density and $v_{\mu}$ is the four velocity of the 
	fluid which satisfies the relation $v^{\mu}v_{\mu}=-1$\,.
	The improved Einstein equations coming from asymptotically safe gravity can be written as
	\begin{eqnarray}
		R_{\mu \nu}-\frac{1}{2}Rg_{\mu \nu}=8\pi G(t)T_{\mu \nu}-\Lambda(t)g_{\mu \nu}\,.
		\label{EE}
	\end{eqnarray}
	In writing down these equations,  a cutoff identification $k=k(t)$ has been made characterized by the cosmic time $t$. From these, we obtain a modified Friedmann equation and a modified continuity equation,
	\begin{align}
		H^{2}=\frac{{\dot{a}}^{2}}{a^2}&=\frac{8\pi}{3}G(t)\rho+\frac{\Lambda(t)}{3} \,  \label{1} \\ \dot{\rho}+3H(p+\rho)&=-\frac{8\pi \rho \dot{G}+ \dot{\Lambda}}{8\pi G(t)}\,.
		\label{mod.cont}
	\end{align}
While this is written for a perfect fluid, we note that $\dot{G}$ and $\dot{\Lambda}$ will appear in the continuity equation for all types of cosmological matter. A similar equation was written in~\cite{lit1} in terms of the scale $k$ for the case where the matter is a scalar field and the cosmological constant enters through the minimum of the scalar potential. In the consistency approach~\cite{Bonanno:2001xi}, it is assumed that the left and right hand sides of Eq.~\eqref{mod.cont} vanish separately. 
The left hand side of Eq.~\eqref{mod.cont} vanishes due to covariant conservation of the energy-momentum tensor $T_{\mu \nu}$\,, which holds when Newton's constant and other coupling constants do not depend on time. The vanishing of the right hand side of Eq.~\eqref{mod.cont} was referred to  as consistency equation. As discussed above, the energy-momentum tensor will not satisfy $\nabla^\mu T_{\mu\nu} = 0\,$ in general if $G, \Lambda$ and  any coupling constant for the matter are time-dependent. Instead we consider the entire right hand side of Eq.~\eqref{EE} to be divergence-free, as required by the contracted Bianchi identity $\nabla^\mu G_{\mu\nu} = 0\,.$ This leads to the modified conservation equation, Eq.~(\ref{mod.cont}).
	
	Using $p(t)=\Omega \rho(t)$\,,  the modified conservation equation can be written as
	\begin{align}
		8\pi \partial_{t}\left[G(t)\rho +\frac{\Lambda(t)}{8\pi}\right]=-24\pi (1+\Omega)HG(t)\rho \,.
		\label{ce}
	\end{align}
	Substituting $G(t)\rho$ from Eq.~\eqref{1} in the above equation, we obtain 
	\begin{align}
		\dot{H}=-\frac{1}{2}(3+3\Omega)\left [ H^{2}-\frac{1}{3}\Lambda(t) \right ]\,.
		\label{3} 
	\end{align}
	In this paper, we wish to solve the above differential equations at late times. To this end, we shall use the long distance perturbative series expansions of $G(k)$ and $\Lambda(k)$ as shown in Eq.~(\ref{G.intro}, \ref{flamda.intro}), suitably converted to a time-varying form.
	The identification of the infrared cutoff for momentum scale $k$ involves expressing $k$ in terms of all scales that are relevant to the problem under consideration. In the case of the FLRW universe, homogeneity and isotropy of spacetime imply that $k$ is a function of the cosmological time only. This in turn implies that the constants $G$ and $\Lambda$ take the form 
	\begin{eqnarray}
		G(t)\equiv G(k=k(t)),\qquad \Lambda(t) \equiv \Lambda(k=k(t))~.
		\label{gk}
	\end{eqnarray}
	The simplest such behaviour at late times is (we will consider some other possibilities later)
	\begin{equation}
		k=\frac{\xi}{t}\,.
		\label{k}
	\end{equation}
	Inserting this expression for $k$ in the series for $G(k)$ and $\Lambda(k)$ 
	(Eqs. (\ref{G}, \ref{flamda}))\,, the time dependent Newton's gravitational constant and cosmological constant in the perturbative or low energy regime is found to be
\begin{align}
G(t) &= G _{0}\left [ 1-\frac{\tilde{\omega}G _{0}}{t^{2}}+\mathcal{O} \left(\frac{t_{Pl}^{4}}{t^{4}} \right ) \right ]\,, \label{gt} \\ 
\Lambda(t) &=\frac{\tilde{\nu }G_{0}}{t^{4}}+\mathcal{O} \left(\frac{t_{Pl}^{6}}{t^{6}} \right )\,,
\label{lambdat}
\end{align}
where $\tilde{\omega} \equiv \omega \xi^{2},~\tilde{\nu} \equiv \nu \xi^{4}$
and $t_{Pl}=\sqrt{G_0}$ is the Planck time in natural units. We see that Newton's constant increases to its classical value $G_0$ at late times, while the cosmological constant approaches zero, which may be appropriate for our universe. We will use these expressions in the next section to investigate the cosmological expansion and entropy generation in our universe. 

	
	\section{Cosmological expansion and entropy generation}\label{Hentropy}
	We shall now solve the differential equation 
	for the Hubble parameter (Eq.\eqref{3}). For this, we set $u(t)=\frac{1}{H(t)}$ and rewrite Eq.~\eqref{3} in terms of $u(t)$ as
	\begin{align}
		\dot{u}=A\left[1-\frac{\Lambda(t) u^{2}(t)}{3}\right]\,
		\label{5}
	\end{align}
	where $A=\frac{(3+3\Omega)}{2}$.
	Integrating this equation from $t=t_{0}$ (present time) to some time $t$, we obtain 
	\begin{align}
		u(t)=u_{0}+A(t-t_{0})-\frac{A}{3}\int_{t_{0}}^{t}\Lambda (t) u^{2}(t)\mathrm{d}t\,
		\label{6} 
	\end{align}
	where $u_{0}\equiv u(t_{0})$. This is a Volterra integral equation of the second kind. We now solve this integral equation iteratively since $\Lambda (t)$ is small at large times (Eq.\eqref{lambdat}). For this, we first approximate our solution by taking 
	\begin{align}
		u(t)\approx u^{(0)}(t)=u_{0}+A(t-t_{0})  \equiv At+B \,
		\label{9}
	\end{align}
	where $B=u_{0}-At_{0}$. Substituting this in the integral on the right hand side of Eq.~\eqref{6}, we obtain to the next approximation
	\begin{align}
		u^{(1)}(t) \simeq u^{(0)}(t)-\frac{A}{3}\int_{t_{0}}^{t}\Lambda (t) \left(u^{(0)}(t)\right)^{2}\mathrm{d}t\,.
		\label{10}
	\end{align}
	Repeating this process by putting $u^{(1)}(t)$ in the integral in Eq.~\eqref{6}, we obtain to the next approximation 
	\begin{align}
		u^{(2)}(t)& \simeq u^{(0)}(t)-\frac{A}{3}\int_{t_{0}}^{t}\Lambda (t) \left(u^{(1)}(t)\right)^{2}\mathrm{d}t \nonumber \\&=u^{(0)}(t)-\frac{A}{3}\int_{t_{0}}^{t}\Lambda (t) \left(u^{(0)}(t)\right)^{2}\mathrm{d}t  + 2\left ( \frac{A}{3} \right )^{2}\int_{t_{0}}^{t}\Lambda (t_{1})\left [\int_{t_{0}}^{t_{1}}\Lambda (t_{2})\left(u^{(0)}(t_{2})\right)^{2}\mathrm{d}t_{2}\right]\mathrm{d}t_{1} \nonumber \\ & \qquad\qquad\qquad\qquad\qquad\qquad\qquad\qquad\qquad\qquad\qquad\qquad\qquad\qquad  +\mathcal{O}(\Lambda^{3}(t))\,.
		\label{11}
	\end{align}
	Substituting $\Lambda(t)$ from Eq.~\eqref{lambdat} in Eq.~\eqref{11} and keeping terms up to $\mathcal{O}(G_{0})$, we get
	\begin{align}
		u^{(1)}(t)\simeq u^{(0)}(t)-\frac{A\tilde{\nu}G_{0}}{3}\left[-\frac{A^{2}}{t}-\frac{AB}{t^{2}}-\frac{B^{2}}{3t^{3}}+\frac{A^{2}}{t_{0}}+\frac{AB}{t_{0}^{2}}+\frac{B^{2}}{3t_{0}^{3}}\right]\,.
		\label{u1}
	\end{align}  
	Defining $C=-\frac{A\tilde{\nu}G_{0}}{3}\left[\frac{A^{2}}{t_{0}}+\frac{AB}{t_{0}^{2}}+\frac{B^{2}}{3t_{0}^{3}}\right]$ for convenience, we get
	\begin{align}
		u(t)\simeq At+B+\frac{A\tilde{\nu}G_{0}}{3}\left[\frac{A^{2}}{t}+\frac{AB}{t^{2}}+\frac{B^{2}}{3t^{3}}\right]+C\,.
		\label{finalu}
	\end{align}
	This now gives a differential equation for the scale factor $a(t)$. This reads   
	\begin{align}
		H=\frac{\dot{a}}{a}&=\frac{1}{At}\left[1+\frac{D}{At}+\frac{\tilde{\nu}G_{0}}{3t}\left(\frac{A^{2}}{t}+\frac{AB}{t^{2}}+\frac{B^{2}}{3t^{3}}\right)\right]^{-1} \nonumber \\&=\frac{1}{At}\left[1-\frac{D}{At}-\frac{\tilde{\nu}G_{0}}{3t}\left(\frac{A^{2}}{t}+\frac{AB}{t^{2}}+\frac{B^{2}}{3t^{3}}\right)+\frac{D^{2}}{A^{2}t^{2}}+\cdots\right]\,
		\label{13}
	\end{align}
	where $D=B+C$. 
	Integrating the above equation, we get upto $\mathcal{O}(\frac{t^{2}_{Pl}}{t^{2}})$
	\begin{align}
		\ln a(t) = \frac{1}{A}\ln t +\frac{D}{A^{2}t}+\frac{A\tilde{\nu }G_{0}}{6 t^2}-\frac{D^{2}}{2A^{3}t^{2}}+\ln \tilde{a} \,
		\label{15}
	\end{align}
	where $\ln \tilde{a}$ is a constant of integration.
	The final form of the scale factor $a(t)$ now reads
	\begin{align}
		a(t) &=\tilde{a} t^{\frac{1}{A}} \exp \left(\frac{D}{A^{2}t}+\frac{A\tilde{\nu }G_{0}}{6 t^2}-\frac{D^{2}}{2A^{3}t^{2}} \right) \nonumber \\   &=\tilde{a} t^{\frac{1}{A}}\left(1+\frac{D}{A^{2}t}+\frac{A\tilde{\nu }G_{0}}{6 t^2}+(1-A)\frac{D^{2}}{2A^{4}t^{2}}+\cdots\right) \,.
		\label{finala}
	\end{align}
	The constant of integration $\tilde{a}$ can be fixed by remembering that $a(t_{0})=1$. This yields 
	\begin{align}
		\tilde{a}=\frac{1}{t_{0}^{\frac{1}{A}}} \left(1+\frac{D}{A^{2}t_{0}}+\frac{A\tilde{\nu }G_{0}}{6 t_{0}^2}+(1-A)\frac{D^{2}}{2A^{4}t_{0}^{2}}+\cdots\right) ^{-1}\,.
		\label{consta}
	\end{align} 
	Eq.~\eqref{finala} gives the quantum correction in the scale factor.
	
	Substituting $\Lambda(t)$ from Eq.~\eqref{lambdat} and the Hubble parameter from Eq.~\eqref{13} in Eq. \eqref{1}, and putting $A=\frac{(3+3\Omega)}{2}$, we get the quantum corrected energy density to be
	\begin{align}
		\rho = \frac{3}{8\pi} \left(\frac{2}{3+3\Omega}\right)^{2}\frac{1}{G_{0}t^{2}} \left[1-\frac{4D}{(3+3\Omega)t}+\frac{\tilde{\omega}G_{0}}{t^2}-\frac{(3+3\Omega)^{2}\tilde{\nu}G_{0}}{4t^{2}}+\frac{4D^{2}}{3(1+\Omega)^{2}t^{2}}+\cdots \right]\,.
		\label{finalrho}
	\end{align}
	Note that the first term is exactly the classical result of energy density in FLRW cosmology. The subsequent terms are the quantum corrections in the energy density  up to $\mathcal{O}(\frac{t_{Pl}^{2}}{t^{2}})$.
	
	Now we proceed to discuss entropy generation in the above scenario in line with the arguments in ~\cite{Bonanno:2007wg, abonanno:2008}.
	From  the modified continuity equation~\eqref{mod.cont}, we get 
	\begin{align}
		\left[\dot{\rho}+3H(p+\rho)\right]\frac{4\pi}{3}R_{0}^{3}a^{3}=\frac{4\pi}{3} R_{0}^{3}\tilde{\mathcal{P}}(t)\,,
		\label{entropy}
	\end{align}
	where we have written
	\begin{align}
		\tilde{\mathcal{P}}(t)=-\left[ \frac{\dot{\Lambda}+8\pi \rho\dot{G}}{8 \pi G} \right ]a^{3}\,,
		\label{Ptil} 
	\end{align} 
	with $R_{0}$ being the radius of the universe at the present time $t_{0}$. Note that  the scale factor here is dimensionless with $a(t)=\frac{R(t)}{R_{0}}$.
	We can rewrite  the above equation as
	\begin{align}
		\frac{\mathrm{d} (\frac{4\pi}{3}\rho R_{0}^{3}a^{3})}{\mathrm{d} t}+p\frac{\mathrm{d} (\frac{4\pi}{3}R_{0}^{3}a^{3})}{\mathrm{d} t}=\frac{4\pi}{3}R_{0}^{3}\tilde{\mathcal{P}}(t) \,. 
		\label{e1}
	\end{align}
	In terms of energy  $U=\frac{4\pi}{3}\rho R_{0}^{3}a^{3}$ and proper volume $V=\frac{4\pi}{3} R_{0}^{3}a^{3}$, the above equation takes the form 
	\begin{align}
		\frac{\mathrm{d}U}{\mathrm{d} t}+p\frac{\mathrm{d} V}{\mathrm{d} t}=\frac{4\pi}{3}R_{0}^{3}\tilde{\mathcal{P}}(t) \,.
		\label{1stlaw}
	\end{align}
	In classical cosmology,  $\tilde{\mathcal{P}}(t)=0$ as $\Lambda$ and $G$ do not depend on time. Therefore we can conclude using the standard thermodynamic relation $\mathrm{d}U+p\mathrm{d}V=T\mathrm{d}S$ and $\tilde{P}(t)=0$, that the entropy does not change during the expansion of the universe. 
	
	Here $\tilde{\mathcal{P}}(t)$ is non-zero because $\Lambda$ and $G$ are functions of time, hence in this case, we have
	\begin{equation}
		T \frac{\mathrm{d}S }{\mathrm{d} t}=\frac{4\pi}{3} R_{0}^{3}\tilde{\mathcal{P}}(t)\,,
		\label{18}
	\end{equation}
	where $S$ is the entropy carried by the perfect fluid inside the comoving volume $V=\frac{4\pi}{3}R_{0}^{3}a^{3}$. 
	From Eqs.~\eqref{1stlaw},\eqref{18}, the change in entropy reads
	\begin{align}
		\frac{\mathrm{d}S}{\mathrm{d} t}=\frac{4\pi}{3}R_{0}^{3}\frac{\tilde{\mathcal{P}}(t)}{T}= \frac{4\pi}{3}R_{0}^{3}\left[\dot{\rho}+3H(\rho+p)\right]\frac{a^{3}}{T}\,.
		\label{19}
	\end{align}
	To calculate the entropy or the change in entropy, we 
	need to know the temperature $T$ as a function of time which is in principle unknown.
	We also require the relation between $p$ and $\rho$ which we have taken to be $p=\Omega \rho$ in our analysis. 
	Assuming radiation dominance, the energy density is taken to be $\rho(t)=\sigma^{4}T^{4}$, where $\sigma \equiv\left(\frac{\pi^{2}n_{\text{eff}}}{30}\right)^{\frac{1}{4}}$, with $n_{\text{eff}}$ being given by $n_{\text{eff}}=n_{b}+\frac{7}{8}n_{f}$, where $n_{b}$  and $n_{f}$ are the bosonic and fermionic massless degrees of freedom respectively. This is in line with the argument in \cite{Bonanno:2007wg} which mentions that significant entropy production takes place only in the radiation dominated universe. Furthermore, 
	since the non-adiabaticity is small, the Stefan-Boltzmann relation among $p$, $\rho$ and $T$ are still valid in the non-equilibrium cosmological scenario with entropy production \cite{Lima:1996, Lima:1997}. 
	
	Using the equation of state $p=\frac{\rho}{3}$, the entropy production rate in terms of the scale factor and energy density is given by
	\begin{align}
		\frac{\mathrm{d}S }{\mathrm{d} t}=\frac{\mathrm{d} }{\mathrm{d} t}\left[\frac{16\pi}{9}\sigma R_{0}^{3}a^{3} \rho^{3/4}\right]\,.
		\label{21}
	\end{align}
	Integrating Eq. \eqref{21}, the final entropy carried by proper volume $V$ reads
	\begin{align}
		S(t)=\frac{16\pi}{9}\sigma R_{0}^{3}a^{3} \rho^{3/4}+S_{c}\,,
		\label{22}
	\end{align}
	where $S_{c}$ is a constant of integration. Note that if $G$ and $\Lambda$
	are independent of time, the radiation energy density obeys $a^3 \rho^{3/4}=constant$ and hence the entropy $S$
	is a constant. However, if $G$ and $\Lambda$ are time dependent, the quantity
	$a^3 \rho^{3/4}$ is a function of time and hence the entropy $S$ is time dependent.
	
	Substituting the expression for the scale factor $a(t)$ and the energy density $\rho(t)$ from  Eqs.~(\ref{finala}, \ref{finalrho}) in the above equation, we obtain the entropy for $\Omega=\frac{1}{3}$ to be
	\begin{align}
		S(t)=\frac{16\pi}{9}\sigma R_{0}^{3}\tilde{a}^{3} \left[ \frac{3}{32\pi G_{0}}\right]^{\frac{3}{4}} \left[ 1+(3\tilde{\omega}-8\tilde{\nu})
		\frac{G_{0}}{4 t^{2}}+\cdots\right]+S_{c} \,.
		\label{23}
	\end{align}
	From the above result, we observe that the leading quantum correction to the entropy for the radiation dominated universe vanishes if  $(3\tilde{\omega}-8\tilde{\nu})=0\,,$ i.e. for $\xi^{2}= \frac{3\omega}{8\nu}$. This is the same value of $\xi$ that one gets from the consistency approach \cite{Bonanno:2001xi}  in which the energy momentum tensor is taken to be covariantly conserved, independent of the 
	scale(time)-dependence of $G$ and $\Lambda$. In such a scenario the entropy production rate is zero and hence the expansion of the universe is adiabatic. 
	We can also obtain an upper bound for the parameter $\xi$ from the positivity of the entropy at all times which imposes $(3\tilde{\omega}-8\tilde{\nu})\geq0$ leading to 
	$\xi \leq \sqrt{\frac{3\omega}{8\nu}}$.

\section{More choices of cut-off}\label{scalea}
In the analysis of the previous section, we chose the energy scale to be inversely proportional to time. However, there can be other choices.	An interesting possibility that was considered in~\cite{Bonanno:2007wg} is to take the energy scale as a function of the Hubble parameter instead of cosmological time. In this section, we obtain solutions for the  scale factor for some power law relations between the energy scale and the Hubble parameter.
\subsection{Cut-off $k=\zeta H^{\frac{1}{4}}$}
In the first instance we choose the cut-off to be  $k=\zeta H^{\frac{1}{4}}$\,,
	%
	%
where $\zeta$ is a dimensionful constant. Substituting this choice of the cut-off $k$ in Eq.~\eqref{flamda} we get
\begin{align}
	\Lambda(t)=\check{\nu}G_{0}H+\cdots\,
		\label{25}
\end{align}
where $\check{\nu}=\zeta^{4} \nu$. Putting Eq.~\eqref{25} in Eq.~\eqref{3}, we obtain a differential equation for $H(t)$\,,
	\begin{align}
		\frac{\dot{H}}{H^{2}}=-\frac{3}{2}(1+\Omega)\left[1-\frac{\check{\nu}G_{0}}{3H}\right] \,.
		\label{26}
	\end{align}
Solving this equation, we get the Hubble parameter as
\begin{align}
		H=\frac{\check{\nu}G_{0}\exp \left(\frac{(1+\Omega)\check{\nu}G_{0}}{2}t\right)} {c\check{\nu}G_{0}+3\exp \left( \frac{(1+\Omega)\check{\nu}G_{0}}{2}t \right)} \,
		\label{30}
\end{align}
where $c$ is an integration constant.
Writing $H(t_{0})=H_{0}$ at present time $t_{0}$, the constant of integration $c$ comes out to be
	\begin{align}
	c=\frac{1}{H_{0}}\left (1 -\frac{3H_0}{\check{\nu}G_{0}}\right) \exp \left( \frac{(1+\Omega)\check{\nu}G_{0}}{2} t_{0}\right)\,.
	\label{constant}
	\end{align}
Substituting the value of $c$ in Eq. \eqref{30} yields
	\begin{align}
H=\frac{\check{\nu}G_{0}\exp \left(\frac{(1+\Omega)\check{\nu}G_{0}}{2}(t-t_{0})\right)} {\left(\frac{\check{\nu}G_{0}}{H_{0}}-3 \right )+3\exp \left( \frac{(1+\Omega)\check{\nu}G_{0}}{2}(t-t_{0}) \right)} ~\,.
		\label{31}
	\end{align}
The scale factor $a(t)$ can now  be obtained from the above equation and reads 
(setting $a(t_0)=1$ at present time $t_0$)
%
%
	%
%
\begin{align}
a(t)=\left\vert 1-\frac{3H_{0}}{\check{\nu}G_{0}}+\frac{3H_{0}}{\check{\nu}G_{0}}\exp \left( \frac{(1+\Omega)\check{\nu}G_{0}}{2}(t-t_{0}) \right) \right\vert^{\frac{2}{3(1+\Omega)}}\,.
\label{cfinala}
\end{align} 
Note that in the classical limit ${{\nu}}\rightarrow 0$, the scale factor in Eq.~(\ref{cfinala}) reduces to 
 \begin{align}
a(t)=\left [ \frac{3(1+\Omega)}{2}H_{0}(t-t_{0})+1 \right ]^{\frac{2}{3(1+\Omega)}}\,.
 \label{classa}
\end{align}
%
Since Eq.~(\ref{1}) implies that $\left(1-\frac{3H_{0}}{\check{\nu}G_{0}} \right)< 0$\, as a consequence of the energy density $\rho$ being positive at all times, we find that the scale factor $a(t)$ has some interesting features. It is finite and slowly varying in the infinite past $t\rightarrow -\infty$\, provided $(1+\Omega)>0$\,,
so at late times an expanding universe emerges out of what was a nearly static universe at very early times. 
Such a universe is called an emergent universe~\cite{Starobinsky, Ellis2004, Murugan2004, Mulryne2005, Mukherjee:2005, Mukherjee:2006ds, abanerjee}.
It is known that one can get an emergent universe if one uses an equation of state of the form $p=A\rho-B\rho^{\frac{1}{2}}$\,, with $A, B>0$\,. However, such an equation of state corresponds to exotic matter. On the other hand, it was shown in~\cite{ijtp2016sg} that it is possible to get an emergent universe using ordinary matter obeying the standard equation of state $p=\Omega \rho$\,, provided matter creation~\cite{prigogine} is assumed. Here we have found yet another example where matter with the usual equation of state $p=\Omega \rho$\, can lead to an emergent universe. This universe has yet another interesting feature. Since $\left(1-\frac{3H_{0}}{\check{\nu}G_{0}} \right) < 0$\,, the scale factor becomes zero at some point of time and then start increasing again exponentially. 
Thus we can call this a \textit{bouncing emergent universe}.
At the bounce the scale factor $a(t)$ vanishes, and therefore $H \rightarrow \infty$, hence it is necessary to include higher quantum corrections in order to fully understand the behaviour there. We shall not attempt to do that in this paper.

\begin{figure}
\includegraphics[scale=.6]{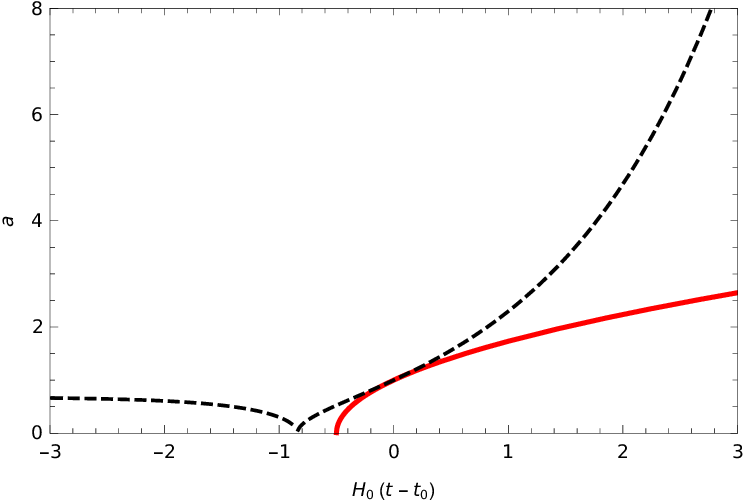}
\caption{Scale factor versus time for a radiation dominated universe. Solid curve: classical scale factor. Dashed curve: quantum corrected scale factor. }
\label{fig1}
\end{figure}

We can put a bound on $\zeta$ by noting that 
away from the classical limit, 
$\zeta^{4} \leqslant \frac{3H_{0}}{\nu G_{0}}$\,. At the critical value $\zeta_c = (\frac{3H_{0}}{\nu G_{0}})^{\frac{1}{4}}$\,, the universe does not have a bounce since the scale factor $a(t) \rightarrow 0$ in the infinite past. The bouncing emergent universe scenario appears whenever $\zeta$ is below its critical value.
Using the observed values $H_{0} \simeq 1.45 \times 10^{-42}$~GeV and $G_{0} \simeq 0.67 \times 10^{-38}$~GeV$^{-2}$~\cite{Ryden}, we find $\zeta_c^{4} = 3.39 \times 10^{-3}~$GeV$^{3}$\,. If we use these values for $H_0\,, G_0\,,$ and also the cosmological constant $\Lambda_0\simeq 4.30 \times 10^{-84}$~GeV$^{2}$\, \cite{Planck}, we find from Eq.~(\ref{25}) that for our universe $\zeta^{4} \simeq 2.32 \times 10^{-3}$~GeV$^{3}$\,.
We have plotted the quantum corrected scale factor $a(t)$ against $H_{0}(t-t_{0})$ for a radiation dominated universe ($\Omega=\frac{1}{3}$)  using this value of $\zeta\,$ in Fig.~\eqref{fig1}.

It is useful to express $H$ as a function of the scale factor $a(t)$\,. For this we note that $\frac{1}{H}\left(\frac{\mathrm{d}H}{\mathrm{d}t}\right)=\left(\frac{\mathrm{d}H}{\mathrm{d}a}\right)\left(\frac{\mathrm{d}a}{\mathrm{d}t}\right) \frac{1}{H}=a\left(\frac{\mathrm{d}H}{\mathrm{d}a}\right)$ and rewrite Eq.~\eqref{26} as
\begin{align}
	a \frac{\mathrm{d} H}{\mathrm{d} a}=-\frac{3+3\Omega}{2}  \left[ H-\frac{\check{\nu}G_{0}}{3}\right] \,.
\label{dH}
\end{align}
Integrating this equation from present time $t_{0}$ to some time $t$, we obtain $H$ in terms of $a$ as
%
%
%
\begin{align}
H=a^{-\frac{3+3\Omega}{2}}H_{0} \left[ 1+\frac{\check{\nu}G_{0}}{3H_{0}}\left(a^{\frac{3+3\Omega}{2}}-1\right)\right]  \,,
\label{finH}
\end{align}
where $H_{0}=H(t_{0})$ and $a(t_{0})=1$\,.
	
In order to calculate the entropy production in this cosmology, we first calculate the energy density $\rho(a)$ using Eq.~\eqref{1}. Inserting the cut-off $k=\zeta H^{\frac{1}{4}}$ into Eq.~\eqref{G} we get
\begin{align}
	G(t)=G_{0}\left[1-\check{\omega}G_{0}H^{\frac{1}{2}}+\cdots\right] \,
		\label{cutg}
\end{align}
where we have defined $\check{\omega}=\zeta^{2}\omega$. 
%
Substituting Eq.~\eqref{cutg} in Eq.~\eqref{1} and keeping terms up to $\mathcal{O}(G_{0})$ inside the parenthesis, we get the energy density in terms of the Hubble parameter as
\begin{align}
\rho(a)=\frac{3H}{8\pi G_{0}}\left[1+\check{\omega}G_{0}H^{\frac{1}{2}}+\cdots\right]\left[H-\frac{\check{\nu}G_{0}}{3}\right] \,.
\label{rhoH}
\end{align}
Using the expression for $H$ in terms of $a$, we can recast the energy density in the form
\begin{eqnarray}
\rho(a)=\frac{3H^{2}_{0}}{8\pi G_{0}}a^{-\frac{(3+3\Omega)}{2}}\left [ 1+\frac{\check{\nu}G_{0}}{3H_{0}}\left ( a^{\frac{3+3\Omega}{2}}-1 \right )\right ] && \left[1+\check{\omega}G_{0}H^{\frac{1}{2}}_{0}a^{-\frac{3+3\Omega}{4}}\left ( 1+\frac{\check{\nu}G_{0}}{3H_{0}}\left ( a^{\frac{3+3\Omega}{2}}-1 \right ) \right )^{\frac{1}{2}}\right] \nonumber \\ &&  \times\left [a^{-\frac{(3+3\Omega)}{2}}\left ( 1+\frac{\check{\nu}G_{0}}{3H_{0}}\left ( a^{\frac{3+3\Omega}{2}}-1 \right ) \right )-\frac{\check{\nu}G_{0}}{3H_{0}}\right ]\,. \nonumber \\
\label{frhoH}
\end{eqnarray}
Now to calculate the entropy produced, we use Eq.~\eqref{22} and set $\Omega=\frac{1}{3}$ (for radiation dominated universe). This gives  
\begin{eqnarray}
S(a)=\frac{16\pi}{9}\sigma R_{0}^{3}&& \left[\frac{3H^{2}_{0}}{8\pi G_{0}}\left ( 1- \frac{\check{\nu}G_{0}}{3H_{0}}\right )\right]^{\frac{3}{4}}\left[\left(1+\frac{\check{\nu}G_{0}}{3H_{0}}\left ( a^{2}-1 \right )\right )\right.\nonumber\\
&&\qquad \left. \times\left ( 1+\frac{\check{\omega}G_{0}H^{\frac{1}{2}}_{0}}{a}\left ( 1+\frac{\check{\nu}G_{0}}{3H_{0}}\left ( a^{2}-1 \right ) \right )^{\frac{1}{2}} \right )\right]^{\frac{3}{4}} +S_{c}.
\label{entropyH}
\end{eqnarray}
From the above expression, we observe that the entropy diverges at late times for the bouncing emergent universe.


\subsection{Cut-off $k=\varepsilon H^{\frac{3}{4}}$}
Of course, for a general cut-off function it is not possible to find closed-form solutions for the scale factor. Let us consider another special case for which a solution can be found, $k=\varepsilon H^{\frac{3}{4}}$, with $\varepsilon$ being a dimensionful constant. 
%
We use this cut-off to find an expression for $\Lambda$ from Eq.~\eqref{flamda},
\begin{align}
		\Lambda(t)=\bar{\nu }G_{0}H^{3}+\cdots \,
		\label{36}
\end{align}
where $\bar{\nu}=\nu \varepsilon^{4}$. Putting $\Lambda(t)$ from  Eq.~\eqref{36} up to order $\mathcal{O}(G_{0})$ in Eq.~\eqref{3}, we get the differential equation for the Hubble parameter to be
\begin{align}
		\dot{H}=-\frac{(3+3\Omega)}{2}\left[H^{2}-\frac{\bar{\nu}G_{0}H^{3}}{3}\right] \,.
		\label{37}
\end{align}
Using $\frac{1}{H}\left(\frac{\mathrm{d}H}{\mathrm{d}t}\right)=
a\left(\frac{\mathrm{d}H}{\mathrm{d}a}\right)$ as before,  we can recast the above equation in the form 
	\begin{align}
		a\left(\frac{\mathrm{d}H}{\mathrm{d}a}\right)=-\frac{(3+3\Omega)}{2} H\left[1-\frac{\bar{\nu}G_{0}H}{3}\right] \,.
		\label{38}
	\end{align}
	Integrating the above equation from the present time $t_{0}$ to some time $t$, we obtain 
	\begin{align}
		\frac{H\left(1-\frac{\bar{\nu}G_{0}}{3}H_{0}\right)}{H_0\left(1-\frac{\bar{\nu}G_{0}}{3}H\right)}=a^{-\frac{(3+3\Omega)}{2}}\,
		\label{40}
	\end{align}
where we have set $a(t_{0})=1$ and written $H(t_{0})= H_{0}$ for the value of the Hubble parameter at the present time. Solving for $H$, we obtain
	\begin{align}
		H=H_{0}a^{-\frac{(3+3\Omega)}{2}}\left[1+\frac{\bar{\nu}G_{0}H_{0}}{3}\left(a^{-\frac{(3+3\Omega)}{2}}-1\right)\right]^{-1}\,.
		\label{f40}
	\end{align}

Solving this equation  gives the scale factor $a(t)$ as 
\begin{align}
\left ( 1-\frac{\bar{\nu}G_{0}H_{0}}{3} \right )\left ( a^{\frac{3+3\Omega}{2}}-1\right )+\frac{(3+3\Omega)}{2}\frac{\bar{\nu}G_{0}H_{0}}{3}\ln a= \frac{(3+3\Omega)}{2} H_{0}\left ( t-t_{0} \right )\,,
\label{asolu}
\end{align}
where the constant of integration has been fixed from the condition $a(t_0)=1$ and we have assumed that $\Omega\neq -1$\,.
The functional dependence of the scale factor $a$ in terms of the cosmological time
$t$ can be obtained iteratively, and up to first order in $G_{0}$ reads
\begin{eqnarray}
a(t)&=&\left[1+\frac{(3+3\Omega)}{2}H_{0}(t-t_{0})
\right.\nonumber\\
&&\left.\qquad+\frac{\bar{\nu}G_{0}H_{0}}{3}\left(\frac{(3+3\Omega)}{2}H_{0}(t-t_{0}) -\ln \left (\frac{(3+3\Omega)}{2}H_{0}(t-t_{0})+1\right ) \right)\right]^{\frac{2}{(3+3\Omega)}}\,.\qquad
\label{ak34a}
\end{eqnarray}
Note that at late times, i.e. as  $t\rightarrow \infty$\,, we find the usual behaviour of classical cosmology, $a\rightarrow \infty$ and $H\rightarrow 0$. 

\begin{figure}
\includegraphics[scale=.6]{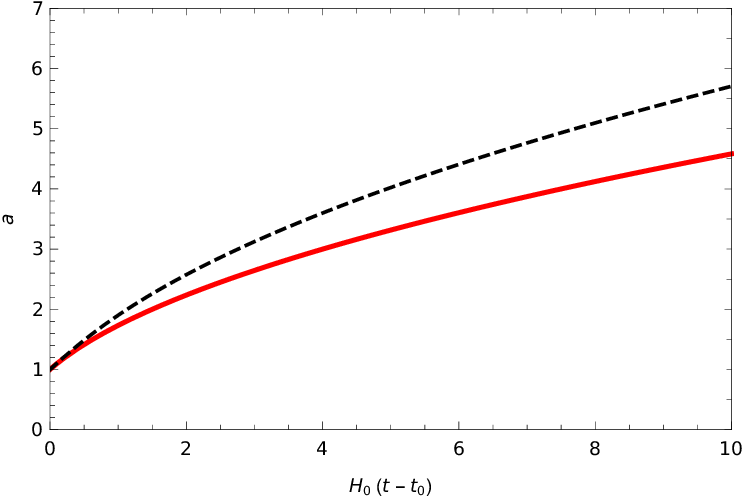}
\caption{Scale factor versus time for a radiation dominated universe. Solid line : classical scale factor. Dashed line : quantum corrected scale factor.}
\label{fig2}
\end{figure}

To get an idea about the energy scale of the cut-off $k$, we estimate the  dimensionful constant $\varepsilon$ appearing in the cut-off using Eq.~\eqref{36}. Using the values of $\Lambda_{0}\,, G_{0}\,,$ and $H_{0}\,$ at the present time, 
we find $\varepsilon^{4} \simeq 11.01 \times 10^{80}$~GeV. It is interesting to note that this energy scale is of the same order of magnitude as the total mass of the universe.  
For these values, $\frac{\bar{\nu}G_{0}H_{0}}{3}$ turns out to be a number of order unity. 
We use this to make a plot of the scale factor $a(t)$ \textit{vs.} $H_{0} (t-t_{0})$ for radiation dominated universe ($\Omega=\frac{1}{3}$) which is displayed  in Fig.\eqref{fig2}.

For the calculation of entropy, we note that using the cut-off $k=\varepsilon H^{\frac{3}{4}}$ in Eq.~\eqref{G}, we get 
\begin{align}
		G(t)=G_{0} \left[1-\bar{\omega}G_{0}H^{\frac{3}{2}}+\cdots\right]
		\label{f40f}
\end{align} 
where $\bar{\omega}=\omega \varepsilon^{2}$. It is clear that $G\rightarrow G_{0}$ at late times. Using this form of $G(t)$, we get the energy density in terms of the Hubble parameter to be 
\begin{align}
\rho(a)=\frac{3H^{2}}{8\pi G_{0}}\left[1+\bar{\omega}G_{0}H^{\frac{3}{2}}+\cdots\right]\left[1-\frac{\bar{\nu}G_{0}}{3}H\right] \,.
		\label{41}
\end{align}
Using the expression for $H$ in Eq.~\eqref{41}, we obtain the energy density in terms of the scale factor $a(t)$ as
\begin{align}
\rho(a)=\frac{3H^{2}_{0}}{8\pi G_{0}}a^{-(3+3\Omega)}\left[1+\bar{\omega}G_{0}H^{\frac{3}{2}}_{0}a^{-\frac{3(3+3\Omega)}{4}}-\frac{\bar{\nu}G_{0}H_{0}}{3}\left(2a^{-\frac{3+3\Omega}{2}}-1\right)+\cdots\right] \,.
\label{42}
\end{align}
The entropy produced can now be calculated in case of the radiation dominated universe by setting $\Omega=\frac{1}{3}$. We find
\begin{align}
S(a)=\frac{16\pi}{9}\sigma R_{0}^{3}\left(\frac{3H^{2}_{0}}{8\pi G_{0}}\right)^{\frac{3}{4}}\left[1+\frac{3\bar{\omega}G_{0}H^{\frac{3}{2}}_{0}}{4a^{3}}-\frac{\bar{\nu}G_{0}H_{0}}{4}\left(\frac{2}{a^{2}}-1\right)+\cdots\right] +S_{c}\,.
\label{43}
\end{align}
The entropy approaches a constant value at late times. 
\subsection{Cut-off $k=\gamma H $}
\begin{figure}
\includegraphics[scale=.6]{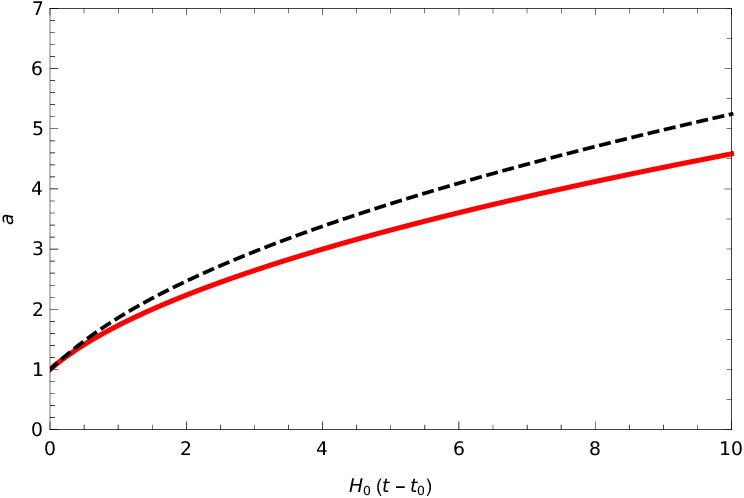}
\caption{Scale factor versus time for a radiation dominated universe. Solid line : classical scale factor. Dashed line : quantum corrected scale factor.}
\label{fig3}
\end{figure}

A cut-off identification suggested in~\cite{Bonanno:2007wg} was $k(t)=\gamma H$\,. For the sake of completeness, we shall now calculate the Hubble parameter and the entropy production for this case.
We now have $\Lambda(t)= \breve{\nu}G_{0} H^{4}+\cdots$ and $G(t)=G_{0}\left[1-\breve{\omega}G_{0}H^{2}+\cdots\right]$ where $\breve{\nu}=\nu \gamma^{4}$ and $\breve{\omega}=\omega \gamma^{2}$. The differential equation \eqref{3} for $H(t)$ for this cut-off reads
%
%
\begin{align}
	a\left(\frac{\mathrm{d}H}{\mathrm{d}a}\right)=-\frac{(3+3\Omega)}{2} H\left[1-\frac{\breve{\nu}G_{0}H^{2}}{3}\right] \,.
	\label{44}
\end{align}
This has the solution
\begin{align}
	H=H_{0}a^{-(3+3\Omega)/2}\left[1-\frac{\breve{\nu}G_{0}H_{0}^{2}}{3}(1-a^{-(3+3\Omega)})\right]^{-\frac{1}{2}}\,.
	\label{45}
\end{align}
Solving this equation and keeping terms up to first order in $G_{0}$ gives the scale factor $a(t)$ as (for $\Omega \neq -1 $) 
\begin{align}
\left ( a^{\frac{3+3\Omega}{2}}-1\right )-\frac{\breve{\nu}G_{0}H_{0}^{2}}{6}\left (a^{\frac{3+3\Omega}{2}}+ a^{-\frac{3+3\Omega}{2}}-2\right ) = \frac{(3+3\Omega)}{2} H_{0}\left ( t-t_{0} \right )\,.
\label{solh3/4}
\end{align}
We perturbatively calculate the functional dependence of the scale factor $a$ in terms of the cosmological time $t$, and up to first order in $G_{0}$ we find
%
%
\begin{eqnarray}
a(t)&=&\left[\frac{(3+3\Omega)}{2}H_{0}(t-t_{0})+1
+\frac{\breve{\nu}G_{0}H_{0}^{2}}{6} \right.\nonumber\\
&&\left. \times \left(\frac{(3+3\Omega)}{2}H_{0}(t-t_{0}) +\frac{1}{\frac{(3+3\Omega)}{2}H_{0}(t-t_{0})+1} -1\right)\right]^{\frac{2}{(3+3\Omega)}}\,.
\label{sola14}
\end{eqnarray}
Once again we find that $a\rightarrow\infty$ as $t\rightarrow\infty$, and therefore $H\rightarrow 0$. We can estimate the dimensionless constant $\gamma$ appearing in the cut-off using Eq.~\eqref{flamda}. As before, we use the observed values of the constants $\Lambda_{0}$, $H_{0}$ and $G_{0}$ at the present time, and thus calculate $\gamma^{4} \simeq 7.59 \times 10^{122}$.  
This in turn makes $\frac{\breve{\nu}G_{0}H_{0}^{2}}{6}$ a number of order unity. Using these numbers, we have plotted the scale factor $a(t)$ against $H_{0} (t-t_{0})$ for the radiation dominated universe,  which is displayed  in Fig.\eqref{fig3}.

Using the expression for $H$ obtained in Eq.~\eqref{45} we calculate the energy density as a function of the scale factor  
\begin{align}
\rho(a)= \frac{3H_{0}^{2}}{8\pi G_{0}}a^{-(3+3\Omega)}\left[1+\frac{2\breve{\nu}G_{0}H_{0}^{2}}{3}\left(\frac{1}{2}-a^{-(3+3\Omega)}\right)+\breve{\omega}G_{0} H_{0}^{2}a^{-(3+3\Omega)}+ \cdots \right]\,.
\label{46}
\end{align}
From this result, we get the entropy in the radiation dominated universe as
\begin{align}
S(a)=\frac{16\pi}{9}\sigma R_{0}^{3}\left(\frac{3H^{2}_{0}}{8\pi G_{0}}\right)^{\frac{3}{4}}\left[1+\frac{3\breve{\omega}G_{0}H^{2}_{0}}{4a^{4}}+\frac{\breve{\nu}G_{0}H_{0}^{2}}{2}\left(\frac{1}{2}-a^{-4}\right)+ \cdots \right] +S_{c}\,.
\label{47}
\end{align}
The entropy is again a constant at late times. 

\section{Summary and Conclusions}
The asymptotic safety approach to quantum gravity results in the running of Newton's gravitational constant $G$ and the cosmological constant $\Lambda$\,. In this paper we have studied the consequence of this scale dependence on the FLRW universe, specifically, the late time behaviour of the scale factor using different choices of the infrared cut-off scale $k$\,.  Our approach is different from an earlier one in that we have considered the conservation of the entire right hand side of Einstein equation.
	
We consider different interpretations of the cut-off scale $k$ in terms of the cosmological time parameter, as is typically done in relating the scale dependent results to cosmological solutions. The first choice of the cut-off scale we take is $k=\frac{\xi}{t}$. Incorporating this into the late time relations for the running of $G$ and $\Lambda$, we solve the differential equations for  the Hubble parameter and the energy density using an iterative approach. The solutions give the quantum corrections for the scale factor and the energy density.  We observe that an $\mathcal{O}(1/t)$ term appears in the quantum correction of the scale factor. This term did not appear in~\cite{Bonanno:2001xi} where the covariant conservation of the energy momentum tensor was imposed separately. Clearly, this behaviour is captured only if the scale (time) dependence of the gravitational and cosmological ``constants'' is taken into account for the conservation of the energy-momentum tensor.
	
We have also calculated the entropy production rate at late times. For this we have assumed a radiation dominated universe and also that the non-adiabaticity is small. Here, we observe that the entropy gets a time dependent quantum correction at  $\mathcal{O}(1/t^2)$. We conclude that the consistency approach of~\cite{Bonanno:2001xi} the vanishing entropy production rate is an artifact of the requirement that $T_{\mu\nu}$ should be conserved independently of the scale (time)-dependence of $G$ and $\Lambda$.
	
Considering the possibility that the cut-off scale may have a simpler functional relationship with the Hubble parameter than with the cosmological time, we looked at three other choices of the cut-off scale, namely, $k=\zeta H^{\frac{1}{4}}$\,, $k=\varepsilon H^{\frac{3}{4}}$\,, and $k = \gamma H$\,. For these also, we have calculated the quantum corrected energy density and entropy in terms of the scale factor for these choices of the cut-off scale. Interestingly, we find that for 	$k=\zeta H^{\frac{1}{4}}$, {{we get a solution for the scale factor which   corresponds to a \textit{bouncing emergent universe}. An emergent universe solution have been obtained earlier in FLRW cosmology using exotic equation of state as well as in cosmologies admitting matter creation. However, in this case the emergent universe like behaviour at the infinite past owes its origin to the running of the constants $G$ and $\Lambda$ as well as the choice of the cut-off scale.}} 
	
Our approach will be useful in situations where the RG flow of the matter sector is included with that of gravity. For example, for electromagnetic radiation one should also include the scale dependence of the fine structure constant which in this picture translates to a time dependence. Time-varying $G$ and $\Lambda$ have been considered even outside asymptotically safe gravity \cite{kall}-\cite{har}, while cosmologies in which the fine structure constant varies with time have been considered in~\cite{barrow}-\cite{uzan}. We expect our approach will be useful in dealing with the time variations of the ``constants'' in such models.

\section*{Acknowledgments} 
RM would like to thank DST-INSPIRE, Govt. of India for financial support.

	
\appendix
\section{Flow of $G$ and $\Lambda$}\label{floweq}
	The starting point is to work with an effective Euclidean action truncated to include only $\sqrt{g}$ and $\sqrt{g}R$ at momentum scale $k$\,,
	\begin{eqnarray}
		\Gamma_{k}[g,\bar{g}]= \frac{1}{16\pi G(k)} \int d^{d}x \sqrt{g}\left\lbrace -R(g)+2{\Lambda}(k)\right\rbrace + S_{gf}[g,\bar{g}]\,,
		\label{EA}
	\end{eqnarray}
where $\bar{g}_{\mu \nu}$ is a background metric and $S_{gf}[g,\bar{g}]$ is a classical background gauge fixing term (see \cite{odi} for a discussion of the gauge dependence of the effective action).
	The flow equation then reads~\cite{Reuter:1996cp, Bonanno:2001xi, Wetterich:1993ya, olauscher:2002ya, Falls:2017lst, dou:1998ya}.  
	\begin{align}
		\partial_{t}\Gamma_{k}[g,\bar{g}]&=\frac{1}{2}\Tr\left[\left(\kappa^{-2}\Gamma_{k}^{(2)}[g,\bar{g}]+\mathcal{R}_{k}^{grav}[\bar{g}]\right)^{-1}\partial_{t}\mathcal{R}_{k}^{grav}[\bar{g}] \right]
		\nonumber \\ &\qquad\qquad 
		- \Tr\left[\left(-M[g,\bar{g}]+\mathcal{R}_{k}^{gh}[\bar{g}]\right)^{-1}\partial_{t}\mathcal{R}_{k}^{gh}[\bar{g}] \right]\,.
		\label{FE} 
	\end{align}
	In this equation
	$\Gamma_{k}^{(2)}[g,\bar{g}]$ is the Hessian of $\Gamma_{k}[g,\bar{g}]$ with respect to $g_{\mu \nu}$\,, $M$ is the Faddeev-Popov ghost operator, while $\mathcal{R}_{k}^{grav}[\bar{g}]$ and $\mathcal{R}_{k}^{gh}[\bar{g}]$ are the infrared regulator functions for gravity and the ghost operator, 
	respectively.
	We have also written $t = \ln k$ and $\kappa = ({32\pi \bar{G}})^{-\frac{1}{2}}$\,, where $\bar{G}$ is the value of $G(k)$ as $k\to \infty$\,.
	
	The infrared regulator function $\mathcal{R}_k$ is chosen so as to suppress modes with momentum below $k$ inside loops. We will take $\mathcal{R}_k$ for both gravity and ghosts to be of the form $\mathcal{R}_{k}(p^2) \propto k^2 R^{(0)}\left(p^2/k^2\right)$ where the function $R^{(0)}(z)$ is smooth and satisfies the conditions $R^{(0)}(0)=1$ and $R^{(0)}(z)\rightarrow 0$ for $z \rightarrow \infty$\,, but is otherwise arbitrary~\cite{Reuter:2001ag, Lauscher:2001ya}.
	When calculating $\Gamma_k$\,, we replace $p^2$ by the quadratic kinetic operator for gravitons or ghosts. Following~\cite{Bonanno:2000ep,Bonanno:2001xi,Reuter:1996cp}, we will take $R^{(0)}(z)$ to be of the form 
	\begin{eqnarray}
		R^{(0)}(z)=z\left[\exp(z)-1\right]^{-1} \,.
		\label{IR}
	\end{eqnarray}
	While other choices for the regulator function are possible, a different choice does not qualitatively change the results~\cite{D.litim:2001, D.litim:2000plb, Reuter:2001ag}.

	Then for $d = 4$, the individual flow equations for $\tilde{g}(k) = k^2 G(k)$ and $\lambda(k) = k^{-2}\Lambda(k) $\,, read
	\begin{align}
		k \partial_{k}\tilde{g} &=(2+\eta_{N})\tilde{g}\,,  \label{ng} \\ 
		k\partial_{k}\lambda &=-(2-\eta_{N})\lambda+\frac{\tilde{g} }{2\pi}\left[10\Phi_{2}^{1}(-2\lambda)-8\Phi_{2}^{1}(0)
		-5\eta_{N}\tilde{\Phi}_{2}^{1}(-2\lambda) \right]\,,
		\label{nlambda}
	\end{align}
	where
	\begin{eqnarray}
		\eta_{N}(\tilde{g} ,\lambda)=\frac{\tilde{g}  B_{1}(\lambda)}{1-\tilde{g}  B_{2}(\lambda)}
		\label{anomalous}
	\end{eqnarray}
	is the 
	anomalous dimension of the operator $\sqrt{g}R$\,. The functions $B_{1}(\lambda)$ and $B_{2}(\lambda)$ are dependent on the regulator function, given by the expressions
	\begin{align}
		B_{1}(\lambda)&\equiv -\frac{1}{3\pi}\left[18\Phi_{2}^{2}(-2\lambda)-5\Phi_{1}^{1}(-2\lambda)+4\Phi_{1}^{1}(0)+6\Phi_{2}^{2}(0) \right]\,,\label{B1} \\ 
		B_{2}(\lambda) &\equiv \frac{1}{6\pi}\left[18\tilde{\Phi}_{2}^{2}(-2\lambda)-5\tilde{\Phi}_{1}^{1}(-2\lambda)\right]\,,
		\label{B2}
	\end{align}
	where functions $\Phi_{n}^{p}(w)$ and $\tilde{\Phi}_{n}^{p}(w)$ are defined as
	\begin{eqnarray}
		\Phi_{n}^{p}(w)&=&\frac{1}{\Gamma(n)}\int_{0}^{\infty}dz z^{n-1}\frac{R^{(0)}(z)-z{R^{(0)}}'(z)}{\left[z+R^{(0)}(z)+w \right]^{p} }\,,
		\label{phi1} \\
		\tilde{\Phi}_{n}^{p}(w)&=&\frac{1}{\Gamma(n)}\int_{0}^{\infty}dz z^{n-1}\frac{R^{(0)}(z)}{\left[z+R^{(0)}(z)+w \right]^{p} }\,.
		\label{phi2}
	\end{eqnarray}
	Recasting Eqs.~(\ref{ng}, \ref{nlambda}) in terms of $G(k)$, $\Lambda(k)$, we finally get
	\begin{align}
		k \partial_{k}G(k) &=\eta_{N}G(k)\,, \label{nG} \\ 
		k\partial_{k}\Lambda(k) &=\eta_{N}\Lambda(k)+\frac{1}{2\pi}k^{4}G(k)\left[10\Phi_{2}^{1}(-2\Lambda(k)/k^2)-8\Phi_{2}^{1}(0) 
		-5\eta_{N}\tilde{\Phi}_{2}^{1}(-2\Lambda(k)/k^2) \right]\,. 
		\label{nLambda}
	\end{align}
	Using the expressions for $B_1$ and $B_2$, we can expand the anomalous dimension of the operator $\sqrt{g}R$ for $d=4$\, in powers of $k^2$\,, 
	\begin{eqnarray}
		\eta_{N} =k^{2} G(k) B_{1}(\Lambda(k)/k^2)\left[ 1+k^2 G(k) B_{2}(\Lambda(k)/k^2)+k^4 G^{2}(k)B^{2}_{2}(\Lambda(k)/k^2)+\cdots\right]
		\,.\qquad
		\label{Anomalous}
	\end{eqnarray}
	%
	%
	Eqs.~(\ref{nG}) and~(\ref{nLambda}) cannot be solved exactly. An iterative procedure 
	is used to find the expressions for $\Lambda$ and $G$ at small $k$\,, starting with $\Lambda=0$ and $\eta_N = 0$\,. 
	Then both the functions $\Phi^p_n(\Lambda/k^2)$ and $\tilde{\Phi}^p_n(\Lambda/k^2)$ are even functions of $k$ and vanish for $k\to 0$ if $p\geq 1\,.$ 
	It follows that both the functions $B_1(\lambda)$ and $ B_2(\lambda) $ and thus also $\eta_N$ are even functions of $k$ at this order of iteration.
	Looking at the equations we see that it follows easily from the iterative procedure 
	that both $\Lambda(k)$ and $G(k)$ can be written as power series of only even powers of $k$\,,
	\begin{align}
		G(k) &= G _{0} \left [ 1-\omega  G_{0} k^{2} +\mathcal{O}(G^{2}_{0}k^{4}) \right ]\,, \label{G} \\ 
		\Lambda(k) & = \nu G_{0} k^{4}  +\mathcal{O}(G^{2}_{0}k^{6})
		\label{flamda}
	\end{align}
	where the constants  $\omega$ and $\nu,~\omega_{1}$ are given by
	\begin{align}
		\omega &= -\frac{1}{2}B_{1}(0)=\frac{1}{6\pi} \left[24 \Phi^{2}_{2}(0)-\Phi^{1}_{1}(0)\right]=\frac{4}{\pi}\left(1-\frac{\pi^{2}}{144}\right)\,, \label{omega} \\ 
		\nu &= \frac{1}{4\pi}\Phi^{1}_{2}(0) = \frac{\zeta(3)}{2\pi}\,, \label{nu} 
	\end{align} 
	since $\Phi^{1}_{1}(0)=\frac{\pi^2}{6}\,, \Phi^{1}_{2}(0) = 2\zeta(3)\,,$ and $ \Phi_{2}^{2}(0)=1$\, for $R_0$ as in Eq.~(\ref{IR}). 
	We note that a different choice of the regulator function $R^{(0)}$ results in a modification of the values of the constants and does not change the form of the power series expansions of $G$ and $\Lambda$\,.

	
\end{document}